\documentclass[twocolumn,showpacs,aps,prl,superscriptaddress]{revtex4}

\usepackage{graphicx}
\usepackage{dcolumn}
\usepackage{amsmath}
\usepackage{epsfig}

\RequirePackage{xspace}
\usepackage{relsize}
\def\BR         {{\ensuremath{\cal B}\xspace}}
\newcommand{\stat}{\ensuremath{\mathrm{(stat)}}\xspace}
\newcommand{\syst}{\ensuremath{\mathrm{(syst)}}\xspace}
\newcommand{\theory}{\ensuremath{\mathrm{(theory)}}\xspace}
\def\piz   {\ensuremath{\pi^0}\xspace}
\def\ellp       {\ensuremath{\ell^+}\xspace}
\def\pim   {\ensuremath{\pi^-}\xspace}
\newcommand{\etapr}{\ensuremath{\eta^{\prime}}\xspace}
\def\Bbar    {\kern 0.18em\overline{\kern -0.18em B}{}\xspace}
\def\BB      {\ensuremath{B\Bbar}\xspace} 
\def\invfb   {\ensuremath{\mbox{\,fb}^{-1}}\xspace}
\def\babar{\mbox{\slshape B\kern-0.1em{\smaller A}\kern-0.1em B\kern-0.1em{\smaller A\kern-0.2em R}}}
\def\pep2{PEP-II}
\def\Y#1S{\ensuremath{\Upsilon{(#1S)}}\xspace}
\def\Dpm     {\ensuremath{D^\pm}\xspace}
\def\Dz      {\ensuremath{D^0}\xspace}
\def\Dp      {\ensuremath{D^+}\xspace}
\newcommand{\gevc}{\ensuremath{{\mathrm{\,Ge\kern -0.1em V\!/}c}}\xspace}
\newcommand{\gevcc}{\ensuremath{{\mathrm{\,Ge\kern -0.1em V\!/}c^2}}\xspace}
\def\jpsi     {\ensuremath{{J\mskip -3mu/\mskip -2mu\psi\mskip 2mu}}\xspace}
\def\pipm  {\ensuremath{\pi^\pm}\xspace}
\def\epem       {\ensuremath{e^+e^-}\xspace}
\def\pip   {\ensuremath{\pi^+}\xspace}
\def\KS    {\ensuremath{K^0_{\scriptscriptstyle S}}\xspace} 
\newcommand{\mevcc}{\ensuremath{{\mathrm{\,Me\kern -0.1em V\!/}c^2}}\xspace}
\newcommand{\mevc}{\ensuremath{{\mathrm{\,Me\kern -0.1em V\!/}c}}\xspace}
\newcommand{\mev}{\ensuremath{\mathrm{\,Me\kern -0.1em V}}\xspace}
\def\ps         {\ensuremath{{\rm \,ps}}\xspace} 
\def\Bz      {\ensuremath{B^0}\xspace}
\def\Bzb     {\ensuremath{\Bbar^0}\xspace}
\def\BzBzb   {\ensuremath{\Bz {\kern -0.16em \Bzb}}\xspace}
\def\Bu      {\ensuremath{B^+}\xspace}
\def\Bub     {\ensuremath{B^-}\xspace}
\def\BpBm    {\ensuremath{\Bu {\kern -0.16em \Bub}}\xspace}
\def\Bp      {\ensuremath{\Bu}\xspace}
\newcommand{\gev}{\ensuremath{\mathrm{\,Ge\kern -0.1em V}}\xspace}
\newcommand{\jprlBase}       {Phys.\ Rev.\ Lett.\xspace}
\newcommand{\jprl}      [1]  {\jprlBase\ {\bf #1}}
\newcommand{\progtp}    [1]  {{Prog.\ Theor.\ Phys.\ {\bf #1}}}
\newcommand{\jplBase}        {Phys.\ Lett.\xspace}
\newcommand{\plb}       [1]  {\jplBase\ B~{\bf #1}}
\newcommand{\nimBaseA}       {Nucl.\ Instrum.\ Methods Phys.\ Res., Sect.\ A\xspace}
\newcommand{\nima}      [1]  {\nimBaseA~{\bf #1}}
\newcommand{\jprBase}        {Phys.\ Rev.\xspace}
\newcommand{\jprd}      [1]  {\jprBase\ D~{\bf #1}}
\newcommand{\jpg}       [1]  {{J.\ Phys.\ {\bf G{\bf #1}}}}
\newcommand{\npps}      [1]  {{Nucl.\ Phys.\ Proc.\ Suppl.\ {\bf #1}}}
\newcommand{\npBase}         {Nucl.\ Phys.\xspace}
\newcommand{\npb}       [1]  {\npBase\ B~{\bf #1}}

\newcommand{\BABARPubYear}    {08}
\newcommand{\BABARPubNumber}  {013}
\newcommand{\SLACPubNumber} {13215}

\def\figurebox#1#2#3{%
    \def\arg{#3}%
    \ifx\arg\empty
    {\hfill\vbox{\hsize#2\hrule\hbox to #2{\vrule\hfill\vbox to #1{\hsize#2\vfill}\vrule}\hrule}\hfill}%
    \else
    {\hfill\epsfbox{#3}\hfill}%
    \fi}

\begin{document}

\preprint{\babar-PUB-\BABARPubYear/\BABARPubNumber} 
\preprint{SLAC-PUB-\SLACPubNumber} 

\begin{flushleft}
\babar-PUB-\BABARPubYear/\BABARPubNumber\\
SLAC-PUB-\SLACPubNumber\\
\end{flushleft}

\title{
{\large \bf Measurements of $B\to\{\pi,\eta,\etapr\}\ell\nu_{\ell}$ Branching Fractions and Determination of $|V_{ub}|$ with Semileptonically Tagged $B$ Mesons }
}

\author{B.~Aubert}
\author{M.~Bona}
\author{Y.~Karyotakis}
\author{J.~P.~Lees}
\author{V.~Poireau}
\author{E.~Prencipe}
\author{X.~Prudent}
\author{V.~Tisserand}
\affiliation{Laboratoire de Physique des Particules, IN2P3/CNRS et Universit\'e de Savoie, F-74941 Annecy-Le-Vieux, France }
\author{J.~Garra~Tico}
\author{E.~Grauges}
\affiliation{Universitat de Barcelona, Facultat de Fisica, Departament ECM, E-08028 Barcelona, Spain }
% ...
\collaboration{The \babar\ Collaboration}
\noaffiliation

\date{\today}

\begin{abstract}
We report measurements of branching fractions for the decays $B\to P\ell\nu_{\ell}$, where $P$ are the pseudoscalar charmless mesons $\pim$, $\piz$, $\eta$ and $\etapr$, based on $348 \invfb$ of data collected with the \babar~detector, using $\Bz$ and $\Bp$ mesons found in the recoil of a second $B$ meson decaying as $B\to D^{(*)}\ell\nu_{\ell}$.  Assuming isospin symmetry, we combine pionic branching fractions to obtain $\BR(\Bz\to\pim\ellp\nu_{\ell}) = (1.54 \pm 0.17_{\stat} \pm 0.09_{\syst}) \times 10^{-4}$; we find $3.2\sigma$ evidence of the decay $\Bp\to\eta\ellp\nu_{\ell}$ and measure its branching fraction to be $(0.64 \pm 0.20_{\stat} \pm 0.03_{\syst}) \times 10^{-4}$, and determine $\BR(\Bp\to\etapr\ellp\nu_{\ell}) < 0.47 \times 10^{-4}$ to $90\%$ confidence level.  Using partial branching fractions for the pionic decays in ranges of the momentum transfer and a recent form factor calculation, we obtain the magnitude of the Cabibbo-Kobayashi-Maskawa matrix element $|V_{ub}| = (4.0 \pm 0.5_{\stat} \pm 0.2_{\syst} {^{+0.7}_{-0.5}}{}_{\theory}) \times 10^{-3}$.
\end{abstract}


\maketitle

The determination of the magnitude of the Cabibbo-Kobayashi-Maskawa matrix \cite{bib_ckm} element $|V_{ub}|$ provides a critical constraint on the Unitarity Triangle; the decay $b\to u\ell\nu_{\ell}$ is a theoretically and experimentally robust means of measuring $|V_{ub}|$.  In the measurements described in this Letter, we reconstruct the $b\to u\ell\nu_{\ell}$ decay exclusively, measuring branching fractions for the processes $\Bz\to\pim\ellp\nu_{\ell}$ \cite{endnote_1} and $\Bp\to\piz\ellp\nu_{\ell}$.  These are selected in the recoil of the semileptonic decay $B \to D^{(*)}\ell\nu_{\ell}$, which provides a measurement complementary to other studies \cite{bib_babartagged, bib_babaruntagged}; this measurement is significantly more precise than previous measurements of its kind \cite{bib_babartagged, bib_belle}.  Additionally, branching fractions for the decays $\Bp\to\eta\ellp\nu_{\ell}$ and $\Bp\to\etapr\ellp\nu_{\ell}$ are measured, which provide potential additional means of determining $|V_{ub}|$ as well as a probe into the dynamics of the $\eta$--$\etapr$ meson system \cite{bib_kim}. 

We use a sample of $383$ million $\BB$ pairs, corresponding to an integrated luminosity of $348 \invfb$ recorded on the $\Y4S$ resonance by the \babar~detector at the \pep2 asymmetric-energy $\epem$ storage rings.  The \babar~detector provides neutral and charged particle reconstruction and charged particle identification, and is described in detail elsewhere \cite{BABARNIM}.  We also use a detailed Monte Carlo simulation (MC) \cite{bib_mc} to estimate signal efficiency and signal and background distributions.

We tag $B$ mesons decaying as $B \to D^{(*)}\ell\nu_{\ell}$ through the full hadronic reconstruction of $\Dpm$ and $\Dz$ mesons; $D^{0}$ mesons are reconstructed through $K^-\pi^+$, $K^-\pip\pip\pim$, $K^-\pip\piz$ and $\KS\pip\pim$ decays, and $D^{+}$ mesons through $K^-\pip\pip$ and $\KS\pip$ decays; $\KS$ candidates are reconstructed as $\KS\to\pip\pim$; and neutral pions are reconstructed as $\piz\to\gamma\gamma$ with the requirement $115 \le m_{\gamma\gamma} \le 150 \mevcc$.  Masses of $D$ candidates are required to be within $2.3\sigma$ of their nominal value, where the mass resolution $\sigma$ ranges between $5.7$ and $19.1\mev/c^2$, depending on the decay channel; we also use a ``sideband'' sample of $D$ candidates with reconstructed mass in a range (typically $4\sigma$ to $7\sigma$) off the appropriate nominal mass.  We require charged daughters of the $D$ candidate to originate from a common vertex.  We reconstruct $D^{*+}$ mesons as $\Dz\pip$ and $\Dp\piz$ and $D^{*0}$ mesons as $\Dz\piz$ and $\Dz\gamma$.  The mass difference between the $D^*$ candidate and its $D$ daughter must be within $3.7\sigma$ of its nominal value; the resolution $\sigma$ of this difference ranges between $0.9$ and $5.7\mev/c^2$, depending on the decay mode.  

Candidate $D^{(*)}$ mesons are paired with tracks identified as leptons with absolute momentum $|\vec{p}_{\ell}| \ge 0.8 \gevc$ \cite{endnote_2}.  If a $D$ candidate (its daughter kaon) is charged, it is required have charge opposite to (same as) that of the corresponding lepton.  The $Y \equiv D^{*}\ell$ system is required to have invariant mass $m_Y \ge 3\gevcc$ and originate from a common vertex.  Photons consistent with originating from bremsstrahlung from this lepton or the decay $D^{(*)}\to D\gamma(\gamma)$ are added to the $Y$ system.  Assuming that the $B \to Y\nu$ decay hypothesis is correct, the angle $\theta_{BY}$ between the directions of the (measured) $Y$ and its parent $B$ is described by
\begin{equation}
\cos\theta_{BY} = \frac{2E_BE_Y - m_B^2 - m_Y^2}{2|\vec{p}_B||\vec{p}_Y|}\mbox{,}
\end{equation}
where $E_B$, $m_B$ and $|\vec{p}_B|$ ($E_Y$, $m_Y$ and $|\vec{p}_Y|$) are the energy, mass and absolute momentum of the $B$ meson ($Y$ system); for the $B$ meson, these are inferred from initial beam energies.  If the $B\to Y\nu$ hypothesis is correct, we have $|\cos\theta_{BY}| \le 1$ up to resolution; because $\cos\theta_{BY}$ is strongly correlated with our discriminating variable $\cos^2\phi_B$, we impose the loose requirement that $|\cos \theta_{BY}| \le 5$.

To suppress background from non-$\BB$ events, we reject events for which the ratio of the second and zeroth Fox-Wolfram moments \cite{bib_foxwolfram} is greater than $0.5$.  We also reject events containing lepton pairs kinematically and geometrically consistent with having originated from the decay of a $\jpsi$ meson.  We reject $D^{(*)}\ell$ candidates for which the event contains any $\KS\to\pip\pim$ candidates not overlapping this $D^{(*)}\ell$ system.  
We require exactly one additional lepton with absolute momentum $|\vec{p}_{\ell}| \ge 0.8 \gevc$ in the event.  If the two leptons are an $\epem$ pair, we require them not to be consistent with originating from $\gamma\to\epem$ conversion.  This second lepton is paired with remaining tracks (assumed to be pions), neutral pions and photons in the event to form $B\to P\ell\nu_{\ell}$ candidates, where $P$ is one of the mesons $\pipm$, $\piz$, $\eta$ or $\etapr$.  For $B\to\pipm\ell\nu_{\ell}$ candidates, the lepton and pion are required to have opposite charge.  $B\to\piz\ell\nu_{\ell}$ candidates are subject to the additional requirement $|\vec{p}_{\piz}| + |\vec{p}_{\ell}| \ge 2.6\gevc$, where $|\vec{p}_{\piz}|$ is the absolute momentum of this $\piz$ candidate.  For $B\to\eta\ell\nu_{\ell}$ candidates, $\eta$ mesons are reconstructed through decays to $\gamma\gamma$, $\pip\pim\piz$ and $\piz\piz\piz$, with invariant mass requirements $500 \le m_{\gamma\gamma} \le 570$, $530 \le m_{\pi\pi\pi} \le 560\mevcc$.  Charged pions from $\eta\to\pip\pim\piz$ decays are required to come from a common vertex; the $\piz$ candidates are required to have absolute laboratory frame momentum greater than $280 \mevc$ ($180\mevc$) when coming from $\pip\pim\piz$ ($\piz\piz\piz$) candidates.  The $\etapr$ meson in $B\to\etapr\ell\nu_{\ell}$ decays is reconstructed through its decay $\etapr\to\eta\pip\pim$ with the $\eta$ candidate selected as above; the additional pions are required to originate from a common vertex, and the $\eta\pip\pim$ system is required to have invariant mass between $920$ and $970\mevcc$.  For $B^{\pm}$ decays ($P = \piz\mbox{,} \eta\mbox{, } \etapr$), the leptons in an event are required to have opposite charge.  

We define the $X$ as a charmless meson $\pipm$, $\piz$, $\eta$ or $\etapr$ and corresponding lepton (including photons consistent with having originated from bremsstrahlung from it); $\theta_{BX}$ is defined analogously to $\theta_{BY}$; we require $|\cos\theta_{BX}| \le 5$.  For each $D^{(*)}\ell$-$P\ell$ candidate, we require that there be no additional tracks in the event and, for hypothesized $\BzBzb$ ($\BpBm$) events, at most $140 \mev$ ($70 \mev$) of neutral energy (i.e., photon candidates) not associated with the $D^{(*)}\ell$ or $P\ell$ candidates.  In the case that more than one $D^{(*)}\ell$-$P\ell$ pair fulfills all requirements for a given event and $P$ mode, the candidate is chosen by smallest $|\cos \theta_{BY}|$, then by largest absolute $P$ momentum.  Signal events containing accepted $D^{(*)}\ell$-$P\ell$ candidates have, on average, between $1.15$ and $1.39$ of them, depending on $P$. 	

Signal yield is extracted independently for each $P$; while we implicitly allow an event to be reconstructed in multiple $P$ modes, we find the induced pairwise statistical correlations between our measured branching fractions to be negligible.  The signal yield is extracted through the quantity $\cos^2 \phi_B$, where $\phi_B$ is the angle between the direction of either $B$ and the plane containing the $X$ and $Y$ momenta:
\begin{equation}
\cos^2\phi_B = \frac{\cos^2\theta_{BY} + 2\cos\gamma\cos\theta_{BY}\cos\theta_{BX} + \cos^2\theta_{BX}}{\sin^2\gamma}\mbox{,}
\end{equation} 
where $\gamma$ is the angle between the $X$ and $Y$ momenta.  For correctly reconstructed signal events, we have $\cos^2\phi_B \leq 1$ up to resolution.

\begin{figure}\begin{center}\includegraphics[width=1.65in]{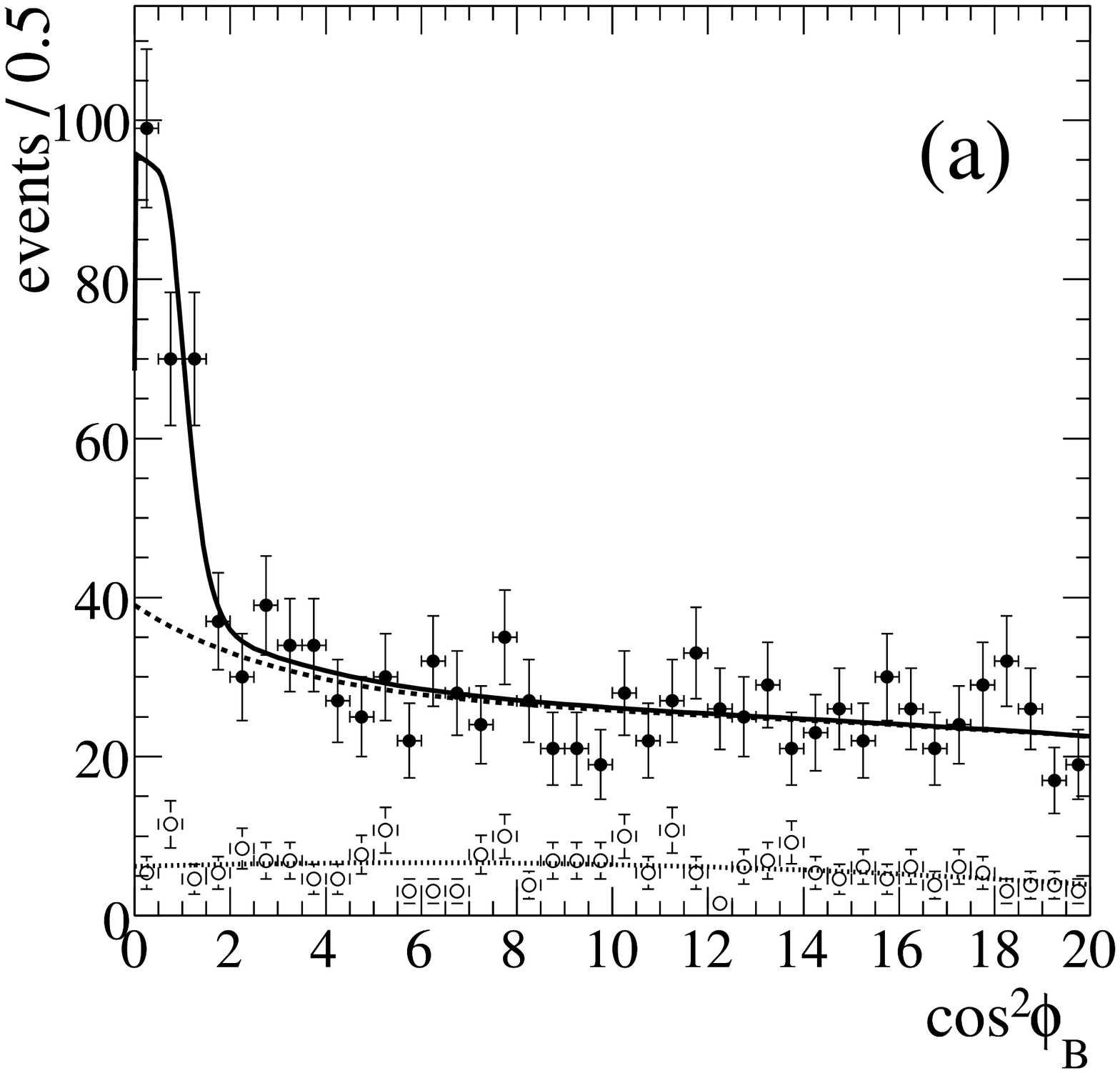}\includegraphics[width=1.65in]{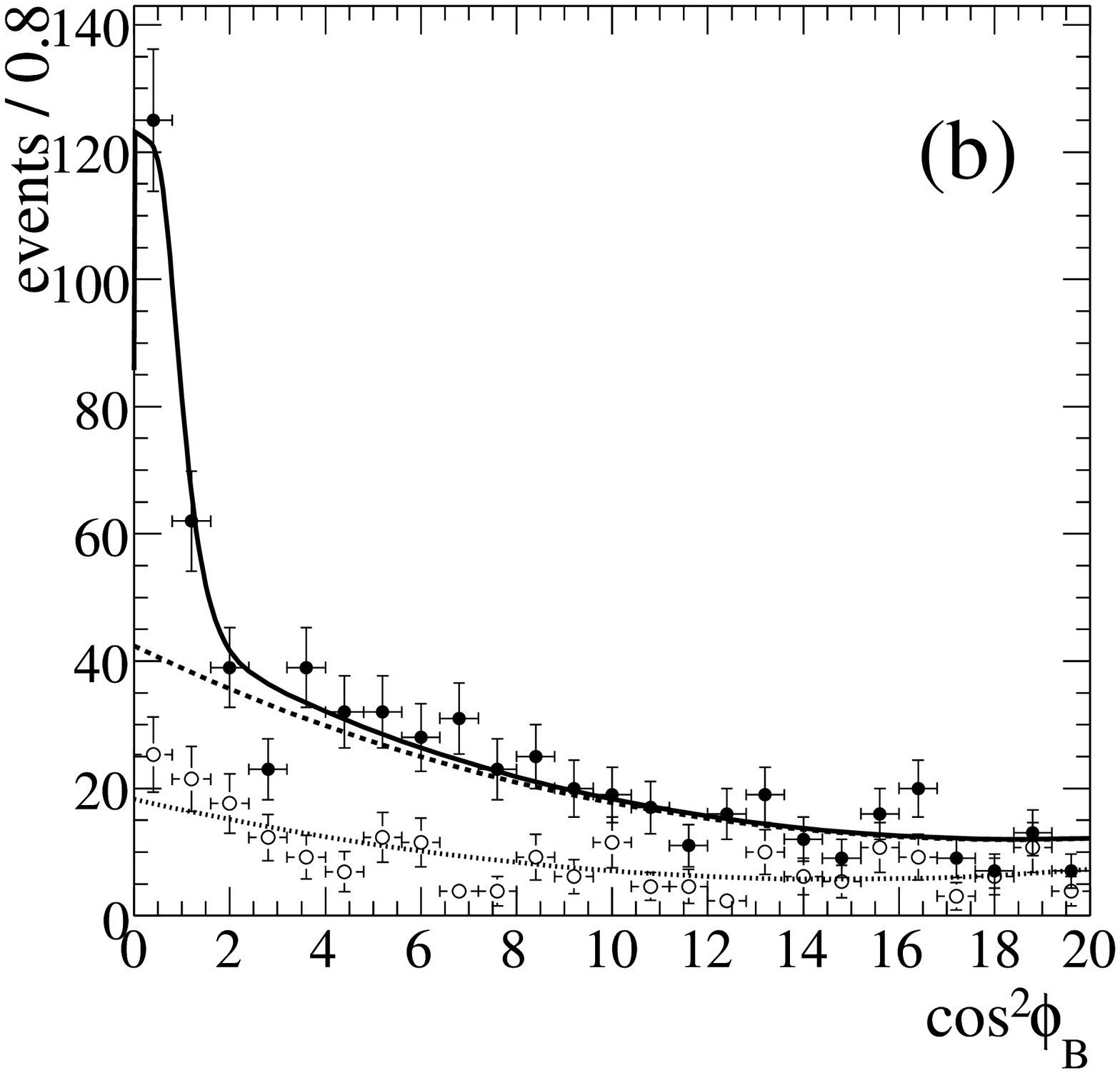}
\includegraphics[width=1.65in]{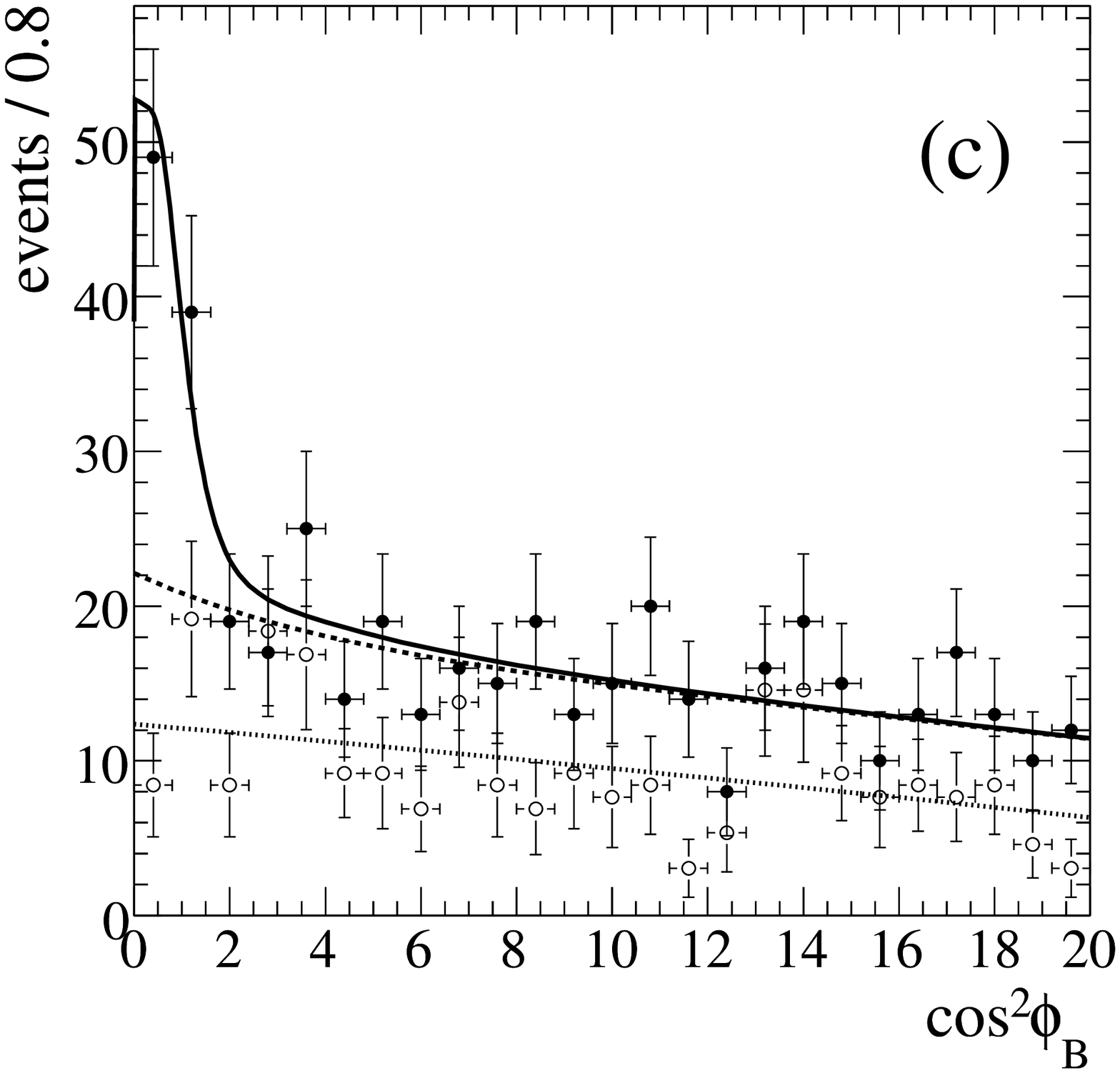}\includegraphics[width=1.65in]{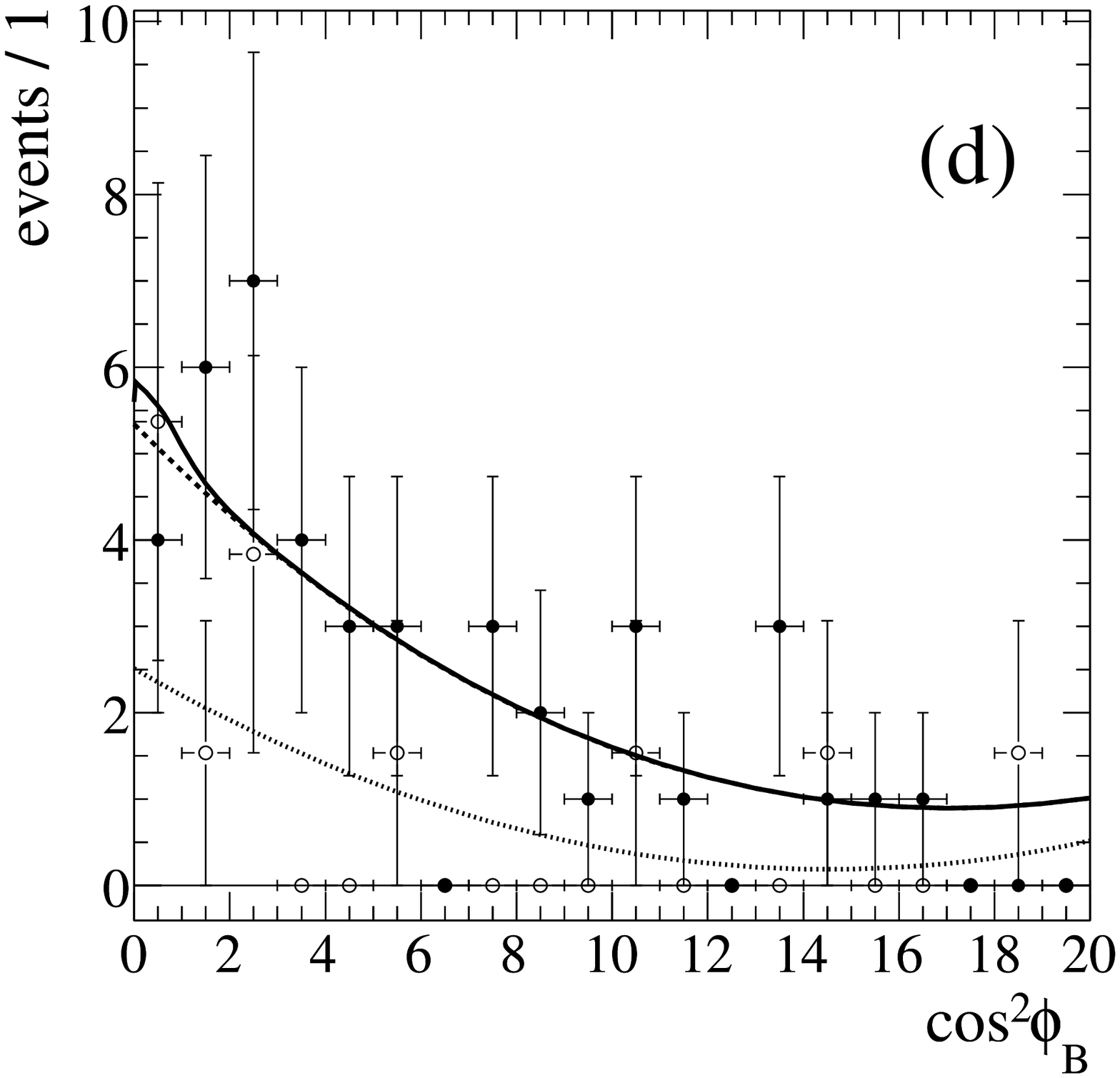}\end{center}\caption{\label{theplot} Distributions of $\cos^2\phi_B$ for $\Bz\to\pim\ellp\nu_{\ell}$ (a), $\Bp\to\piz\ellp\nu_{\ell}$ (b), $\Bp\to\eta\ellp\nu_{\ell}$ (c) and $\Bp\to\etapr\ellp\nu_{\ell}$ (d) candidates; filled and hollow circles represent $D$ mass peak and sideband data, respectively.  The curves are stacked fit results for ``cmb'' (dotted), ``bg'' (dashed) and ``sig'' (solid) PDFs, as defined in the text.  The fits are performed in bins of $q^2$ but are here shown in the full $q^2$ range.}
\end{figure}

For a $B \to P\ell\nu_{\ell}$ decay, $q^2$ is defined as the squared invariant mass of the lepton-neutrino system, and is calculated in the approximation that the $B$ is at rest, i.e., $q^2 = (m_B - E_P)^2 - |\vec{p}_P|^2$, where $E_P$ and $\vec{p}_P$ are, respectively, the energy and momentum of the $P$ meson.  The data are divided into three bins: $q^2 < 8$, $8 \le q^2 < 16$ and $q^2 \ge 16 \gev^2/c^2$, in each of which the yield is extracted separately, except in the $\Bp\to\etapr\ellp\nu_{\ell}$ mode, in which, due to a lower reconstruction efficiency, the yield is measured in a $q^2 < 16 \gev^2/c^2$ bin and over the full $q^2$ range.  The data is described as a sum of three contributions, $dN/d\cos^2\phi_B = N_{\mathrm{sig}}\mathcal{P}_{\mathrm{sig}} + N_{\mathrm{bg}}\mathcal{P}_{\mathrm{bg}} + N_{\mathrm{cmb}}\mathcal{P}_{\mathrm{cmb}}$, where these $N_i$ and $\mathcal{P}_i$ are the yield and probability density functions (PDF) of: signal (``sig''), background with correctly reconstructed $D^{0,\pm}$ mesons (``bg'') and backgrounds with combinatoric $D^{0,\pm}$ candidates (``cmb'').  The signal PDF, $\mathcal{P}_{\mathrm{sig}}$, is modeled as a threshold function (constant between zero and unity, vanishing elsewhere) with finite resolution and an exponential tail (four parameters).  The correct $D$ background PDF, $\mathcal{P}_{\mathrm{bg}}$, is modeled as an exponential with a nonnegative constant term (two parameters); the combinatoric $D$ background, $\mathcal{P}_{\mathrm{cmb}}$, is modeled by a second order polynomial (two parameters).  These eight PDF shape parameters and the $\mathcal{P}_i$ are determined via simultaneous unbinned maximum likelihood fit (see Figure \ref{theplot}) of $dN/d\cos^2\phi_B$ to the data, $\mathcal{P}_{\mathrm{sig}}$ to MC signal events, $\mathcal{P}_{\mathrm{bg}}$ to MC background events (with correctly identified $D^{0,\pm}$ mesons) and $\mathcal{P}_{\mathrm{cmb}}$ to the sideband sample.  The combinatoric yield $N_{\mathrm{cmb}}$ is further constrained, up to statistical accuracy, by the number of events in the sideband sample.  Total signal yields are found to be $150 \pm 22$, $134\pm 20$, $55\pm 15$ and $0.6 \pm 3.9$ events for $\pipm\ell\nu_{\ell}$, $\piz\ell\nu_{\ell}$, $\eta\ell\nu_{\ell}$ and $\etapr\ell\nu_{\ell}$ respectively.

The $B\to D^{(*)}\ell\nu_{\ell}$ reconstruction efficiency is determined via an analogous $\cos^2\phi_B$ study on ``double tag'' events, i.e., events reconstructed as $\BB$ with both $B$ mesons decaying as $B\to D^{(*)}\ell\nu_{\ell}$.  The $B\to P\ell\nu_{\ell}$ reconstruction efficiency for each $q^2$ bin is determined from the MC signal sample, as are bin-to-bin migrations due to the finite $q^2$ resolution, which are small ($<9\%$).  Overall efficiencies, including branching fractions and reconstruction efficiency of the recoil $B$, are found, in units of $10^{-3}$, to be $1.4$, $1.8$, $1.1$ and $0.22$ for $B\to\pipm\ell\nu_{\ell}$, $B\to\piz\ell\nu_{\ell}$, $B\to\eta\ell\nu_{\ell}$ and $B\to\etapr\ell\nu_{\ell}$ respectively.

Systematic uncertainties associated with physics modeling are evaluated by determining the change in the measured branching fraction after varying independently in MC within current knowledge: $B\to\{\rho,\omega\}\ell\nu_{\ell}$ branching fractions, $B\to\pi^{\pm,0}\ell\nu_{\ell}$ branching fractions, $B\to\eta^{(\prime)}\ell\nu_{\ell}$ branching fractions, the total $B$ charmless semileptonic decay branching fraction, the $B$ charmless semileptonic decay spectrum \cite{bib_buchmuller}, $B\to P\ell\nu_{\ell}$ form factors (comparing the model by Ball and Zwicky \cite{bib_ball} to that of Scora and Isgur \cite{bib_scora}) and several $B\to D\ell\nu_{\ell}$ branching fractions; the largest is found to have an effect four times smaller than the statistical uncertainty.  We also apply uncertainties derived from those on $\eta$ and $\etapr$ decay branching fractions.  

We estimate the systematic uncertainty associated with the accuracy of $\BB$ background simulation by comparing the $\cos^2\phi_B$ distributions in signal-depleted data and MC samples.  From study of $37 \invfb$ of $\epem$ collisions $40\mev$ below the $\Y4S$ resonance, we determine that there is no contribution from non-$\BB$ events to the signal; the precision to which this can be determined is also taken as a systemic uncertainty.  

Final state radiation in $\Bz\to\pim\ellp\nu_{\ell}$ decays is determined, from simulation, to cause $q^2$ bin migrations no greater than $1.2\%$, which is conservatively applied as a systematic uncertainty, as well as to the other branching fractions.  We apply a $0.59\%$ ($1.7\%$) systematic uncertainty for $\BzBzb$ ($\BpBm$) decays associated with the assumption that double tag events can be used to estimate the single tag efficiency reliably.

\begin{table*}
\caption{\label{table_result} Partial and total branching fractions, in units of $10^{-4}$, for each decay channel; the first uncertainty given is statistical, the second is systematic.  Ranges for $q^2$ are given in $\gev^2/c^2$.  In the bottom row is the result from combining $\Bz\to\pim\ellp\nu$ and $\Bp\to\piz\ellp\nu$ branching fractions.}
\begin{center}
\begin{tabular}{lcclr@{}ll}
\hline\hline
 & $q^2 < 8$ & $8 \le q^2 < 16$ & \multicolumn{1}{c}{$q^2 \ge 16$} & \multicolumn{2}{c}{$q^2 < 16$} & \multicolumn{1}{c}{total}\\
\hline
$\Bz\to\pim\ellp\nu$ & $0.59 \pm 0.12 \pm 0.03$ & $0.34 \pm 0.11 \pm 0.02$ & $0.46 \pm 0.14 \pm 0.03$ & & $0.92 \pm 0.16 \pm 0.05$ & $1.38 \pm 0.21 \pm 0.07$\\
$\Bp\to\piz\ellp\nu$ & $0.43 \pm 0.09 \pm 0.02$ & $0.29 \pm 0.08 \pm 0.03$ & $0.24 \pm 0.09 \pm 0.03$ & & $0.73 \pm 0.12 \pm 0.05$ & $0.96 \pm 0.15 \pm 0.07$\\
$\Bp\to\eta\ellp\nu$ & $0.28 \pm 0.10 \pm 0.01$ & $0.16 \pm 0.11 \pm 0.01$ & $0.21 \pm 0.13 ^{+0.02}_{-0.01}$ & & $0.43 \pm 0.15 \pm 0.02$ & $0.64 \pm 0.20 \pm 0.03$\\
$\Bp\to\etapr\ellp\nu$ &  - &  - & \multicolumn{1}{c}{-} & $-$ & $0.05 \pm 0.22 ^{+0.04}_{-0.06}$ & $0.04 \pm 0.22 ^{+0.05}_{-0.02}$ \\

\hline

$\Bz\to\pim\ellp\nu$ (combined) & $0.67\pm0.10\pm0.03$ & $0.43\pm0.09\pm0.03$ & $0.46\pm0.11\pm0.04$ & & $1.08\pm0.13{^{+0.05}_{-0.06}}$ & $1.54\pm0.17\pm0.09$\\
 
\hline\hline
\end{tabular}
\end{center}
\end{table*}

As double tag events are used to determine the $D^{(*)}\ell\nu_{\ell}$ reconstruction efficiency, detector simulation uncertainties are applied only to particles on the $P\ell$ side: $0.36\%$ per track, $3\%$ per $\piz$, $2\%$ ($3\%$) per electron (muon).  There is a $1.1\%$ systematic uncertainty from counting $\BB$ pairs \cite{bib_bcounting}, and a $1.4\%$ systematic uncertainty from the $\Y4S\to\BzBzb$ fraction \cite{bib_pdg}.  Measured branching fractions and associated uncertainties are given in Table \ref{table_result}.  Quoted statistical uncertainties are due to the finite size of data and MC samples.  We combine $\Bz\to\pim\ellp\nu_{\ell}$ and $\Bp\to\piz\ellp\nu_{\ell}$ branching fractions using the isospin relation $\Gamma(\Bz\to\pim\ellp\nu_{\ell}) = 2\Gamma(\Bp\to\piz\ellp\nu_{\ell})$ and the lifetime ratio $\tau_{\Bp}/\tau_{\Bz} = 1.071 \pm 0.009$ \cite{bib_pdg}.  The significance of the $\Bp\to\eta\ellp\nu_{\ell}$ signal is $3.2\sigma$.

A Bayesian $90\%$ confidence limit $\BR(\Bp\to\etapr\ellp\nu_{\ell}) < 0.47 \times 10^{-4}$ is determined, assuming a flat prior in the physical (nonnegative branching fraction) region, via the integral of the likelihood function from the signal extraction, smeared by a Gaussian resolution function with varying width representing all other sources of uncertainty.  We also determine the partial branching fraction $\Delta\BR(\Bp\to\etapr\ellp\nu_{\ell}) < 0.37 \times 10^{-4}$ for $q^2<16\gev^2/c^2$ and the ratio $\BR(\Bp\to\etapr\ellp\nu_{\ell})/\BR(\Bp\to\eta\ellp\nu_{\ell}) < 0.57$ with $90\%$ confidence level, the latter of particular importance in constraining the dynamics of the $\eta$-$\etapr$ system \cite{bib_kim}.  These are in disagreement with a recently published result \cite{bib_adam}.

\begin{table}
\caption{\label{table_vub} Values of $|V_{ub}|$ derived using branching fractions measured in this Letter and various form factor calculations.  Range for $q^2$ is stated in $\gev^2/c^2$, reduced decay rate in $\mathrm{ps}^{-1}$.  The given uncertainties on $|V_{ub}|$ are, respectively, statistical, systematic and due to uncertainties in form factor calculation.}
\begin{center}
\begin{tabular}{lccc}
\hline\hline
 & $q^2$ & $\Delta\zeta$ & $|V_{ub}|$ ($10^{-3}$) \\
\hline
Ball \& Zwicky \cite{bib_ball} & $<16$ & $5.44 \pm 1.43$ & $3.6 \pm 0.2 \pm 0.1 ^{+0.6}_{-0.4}$\\
Gulez et al. \cite{bib_gulez}  & $>16$ & $2.07 \pm 0.57$ & $3.8 \pm 0.4 \pm 0.2 ^{+0.7}_{-0.4}$\\
Okamoto et al. \cite{bib_okamoto} & $>16$ & $1.83 \pm 0.50$ & $4.0 \pm 0.5 \pm 0.2 ^{+0.7}_{-0.5}$\\
Abada et al. \cite{bib_abada} & $>16$ & $1.80 \pm 0.86$ & $4.1 \pm 0.5 \pm 0.2 ^{+1.6}_{-0.7}$\\
\hline\hline
\end{tabular}
\end{center}
\end{table}

Extraction of $|V_{ub}|$ from the measured $B\to\pi\ell\nu_{\ell}$ branching fractions $\Delta\BR$ proceeds through the relation $|V_{ub}| = \sqrt{\Delta\BR / (\tau_{\Bz} \Delta\zeta)}$, with $\tau_{\Bz} = 1.530 \pm 0.009 \ps^{-1}$ the $\Bz$ meson lifetime \cite{bib_pdg} and $\Delta\zeta$ the calculated reduced (i.e., appropriately normalized) decay rate over the corresponding $q^2$ range, which depends on the decay form factor $f^{\pi}_+$.  Several form factor calculations are available, including one using light-cone sum rules \cite{bib_ball} and various lattice QCD methods \cite{bib_gulez, bib_okamoto, bib_abada}.  Results are given in Table \ref{table_vub}.  The branching fractions $\BR(B\to\eta^{(\prime)}\ell\nu_{\ell})$ will provide additional means of determining $|V_{ub}|$ as accurate calculations of $f^{\eta^{(\prime)}}_+$ become available.

In conclusion, we have measured the branching fractions for $B\to P\ell\nu_{\ell}$, where $P$ are charmless pseudoscalar mesons, as a function of the squared momentum transfer $q^2$.  We report the total branching fractions, the third with a significance of $3.2\sigma$:
\begin{align}
\BR(\Bz\to\pim\ellp\nu_{\ell}) &= (1.38 \pm 0.21 \pm 0.07) \times 10^{-4}\mbox{,}\\
\BR(\Bp\to\piz\ellp\nu_{\ell}) &= (0.96 \pm 0.15 \pm 0.07) \times 10^{-4}\mbox{,}\\
\BR(\Bp\to\eta\ellp\nu_{\ell}) &= (0.64 \pm 0.20 \pm 0.30) \times 10^{-4}\mbox{,}
\end{align}
with the first uncertainty statistical and the second systematic, and, to $90\%$ confidence level,
\begin{equation}
\BR(\Bp\to\etapr\ellp\nu_{\ell}) < 0.47 \times 10^{-4}\mbox{.}
\end{equation} We combine the pionic branching fractions to obtain $\BR(\Bz\to\pim\ellp\nu_{\ell}) = (1.54 \pm 0.17 \pm 0.09) \times 10^{-4}$, among the most precise measurements of this branching fraction available.  We use the partial branching fractions to extract $|V_{ub}|$, using a variety of form factor calculations, and obtain values ranging from $3.6 \times 10^{-3}$ to $4.1 \times 10^{-3}$.  The pionic branching fraction measurements represent a roughly $30\%$ improvement over a previous \babar~measurement in this channel \cite{bib_babartagged}, and is statistically independent of similar measurements in other channels \cite{bib_babartagged, bib_babaruntagged}.

We are grateful for the excellent luminosity and machine conditions
provided by our \pep2\ colleagues, 
and for the substantial dedicated effort from
the computing organizations that support \babar.
The collaborating institutions wish to thank 
SLAC for its support and kind hospitality. 
This work is supported by
DOE
and NSF (USA),
NSERC (Canada),
CEA and
CNRS-IN2P3
(France),
BMBF and DFG
(Germany),
INFN (Italy),
FOM (The Netherlands),
NFR (Norway),
MES (Russia),
MEC (Spain), and
STFC (United Kingdom). 
Individuals have received support from the
Marie Curie EIF (European Union) and
the A.~P.~Sloan Foundation.


\begin{thebibliography}{99}
\bibitem{endnote_1} Here, $\ell$ or ``lepton'' means electron or muon; charge conjugate states are assumed throughout this Letter.
\bibitem{endnote_2} Unless otherwise noted, all quantities are given in the $\Y4S$ center-of-mass frame.
\end{thebibliography}
\end{document}